\documentclass[aps,pra,amsmath,reprint,floatfix]{revtex4-1}

\usepackage{microtype} 
\usepackage{amssymb}
\usepackage{graphicx}
\usepackage{bbm}
\usepackage{bm}

\RequirePackage[
  hyperindex,colorlinks,bookmarksnumbered,
  plainpages=true,pdfstartview=FitH]{hyperref}
\hypersetup{linkcolor=blue,urlcolor=blue,citecolor=blue} 
\usepackage{hyperref}
\usepackage[all]{hypcap}

\newcommand{\ED}{.}
\newcommand{\EC}{,}

\newcommand{\ER}[1]{Eq.~(\ref{#1})}
\newcommand{\EsR}[1]{Eqs.~(\ref{#1})}
\newcommand{\ERn}[1]{(\ref{#1})}
\renewcommand{\FR}[1]{Fig.~\ref{#1}}
\newcommand{\FsR}[1]{Figs.~\ref{#1}}
\newcommand{\FRn}[1]{\ref{#1}}
\newcommand{\SR}[1]{Sec.~\ref{#1}}
\newcommand{\AR}[1]{App.~\ref{#1}}

\newcommand{\N}{\mathcal{N}}
\newcommand{\p}{^{\prime}}
\newcommand{\pp}{^{\phantom\prime}}
\newcommand{\vpp}{^{\vphantom\prime}}
\newcommand{\pd}{\vphantom{\dot{G}}} 
\newcommand{\pss}{^{\vphantom{(6)}}} 

\renewcommand{\vec}[1]{\bm{#1}}

\allowdisplaybreaks

\begin{document}

\title{Counting Feynman diagrams via many-body relations}
\author{Fabian B.~Kugler}
\affiliation{Physics Department, Arnold Sommerfeld Center for Theoretical Physics, and Center for NanoScience, Ludwig-Maximilians-Universit\"at M\"unchen, Theresienstr.~37, 80333 Munich, Germany}

\date{\today}

\begin{abstract}
We present an iterative algorithm
to count Feynman diagrams
via many-body relations.
The algorithm allows us to count the number of diagrams
of the exact solution for
the general fermionic many-body problem
at each order in the interaction.
Further, we apply it to different
parquet-type approximations and consider spin-resolved
diagrams in the Hubbard model.
Low-order results and asymptotics are explicitly discussed for
various vertex functions and different two-particle channels.
The algorithm can easily be implemented
and generalized to many-body relations
of different forms and levels of approximation.
\end{abstract}

\maketitle

\section{Introduction}
In the study of many-particle systems,
Feynman diagrams are a ubiquitous, powerful tool
to perform and organize
perturbation series as well as partial resummations thereof.
To gain intuition about 
the strength of a diagrammatic
resummation or to compare different 
variants of resummation,
it can be useful to count the number of 
diagrams involved, ideally for several
kinds of vertex functions.
Moreover, the factorial growth in the number of diagrams
with the interaction order is often linked with the
nonconvergent, asymptotic nature of (bare) perturbation series
\cite{Negele2008}.
The asymptotic number of diagrams 
generated by approximate solutions
is therefore of particular interest.
In this paper, we present an algorithm to count
the number of Feynman diagrams inherent in 
many-body integral equations.
Its iterative structure
allows us to numerically access 
arbitrarily large interaction orders
and to gain analytical insights about the asymptotic behavior.
In \SR{sec:relations}, we recapitulate typical many-body relations
as a basis for the algorithm.
The algorithm is explained in \SR{sec:diagr_count},
where some general parts of the discussion follow
Ref.\ \onlinecite{Kugler2017a} quite closely;
some of the ideas have also been formulated by Smith \cite{Smith1992}.
In \SR{sec:results}, we use the algorithm 
to count the exact number of bare and skeleton diagrams of the general many-body problem
for various vertex functions and to discuss their asymptotics.
Subsequently, 
we consider parquet-type approximations as examples for approximate solutions,
and we focus on the Hubbard model to discuss spin-resolved diagrams.
Finally, we present our conclusions in \SR{sec:conclusions}.
\section{Many-body relations}
\label{sec:relations}
A general theory of interacting fermions is defined by the action
\begin{align}
S & = - \sum_{x\p, x} \bar{c}_{x\p} \big( G_0^{-1} \big)_{x\p, x} c_{x} 
- \tfrac{1}{4} 
\!\!\!\! \sum_{x\p,x,y\p,y} \!\!\!\!
\Gamma^{(4)}_{0;x\p,y\p;x,y} \bar{c}_{x\p} \bar{c}_{y\p} c_{y} c_{x}
\EC
\raisetag{6pt}
\label{eq:general_theory}
\end{align}
where $G_0$ is the bare propagator, 
$\Gamma^{(4)}_0$ the bare four-point vertex, which is
antisymmetric in its first and last two arguments,
and $x$ denotes all quantum numbers of the Grassmann field $c_x$.
If we choose, e.g., Matsubara frequency, momentum, and spin, with
$x = (i\omega, \vec{k}, \sigma) = (k, \sigma)$, and consider
a translationally invariant system with interaction $U_{|\vec{k}|}$,
the bare quantities read
\begin{subequations}
\begin{align}
G_{0;x\p,x} 
& 
\overset{\textrm{e.g.}}{=}
G_{0;k,\sigma} \delta_{k\p,k\pp} \delta_{\sigma\p,\sigma\pp}
\EC \\
- \Gamma^{(4)}_{0;x_1\p,x_2\p;x_1\pp,x_2\pp} 
& 
\overset{\textrm{e.g.}}{=}
(
U_{|\vec{k}_1\p-\vec{k}_1\pp|} \delta_{\sigma_1\p,\sigma_1\pp} 
\delta_{\sigma_2\p,\sigma_2\pp}
\nonumber \\ & \, \ -
U_{|\vec{k}_1\p-\vec{k}_2\pp|} \delta_{\sigma_1\p,\sigma_2\pp} 
\delta_{\sigma_2\p,\sigma_1\pp}
) \, 
\delta_{k_1\p+k_2\p,k_1\pp+k_2\pp}
\ED
\label{eq:vertex_U}
\end{align}
\end{subequations}
Interested in one- and two-particle correlations,
the many-body theory is usually focused on
the full propagator $G$ with self-energy $\Sigma$
and the full one-particle-irreducible (1PI) four-point vertex $\Gamma^{(4)}$, which
can be decomposed into two-particle-irreducible vertices $I_r$ in 
different two-particle channels $r \in \{a,p,t\}$ (see below).
The quantities $G$, $\Sigma$, $\Gamma^{(4)}$ 
are related by the exact and closed set
of equations \cite{Hedin1965,Bickers2004,Held2011,Kugler2018}
\begin{subequations}
\begin{align}
G & = G_0 + G_0 \cdot \Sigma \cdot G
\EC \label{eq:dyson} \\
\Sigma & = - \Gamma^{(4)}_0 \circ G 
- \tfrac{1}{2} \Gamma^{(4)}_0 \circ G \circ G \circ G \circ \Gamma^{(4)}
\EC \label{eq:SchwingerDyson} \\
\Gamma^{(4)} & = I_{t} - I_{t} \circ G \circ G \circ \Gamma^{(4)} \EC \quad
I_{t} = - \frac{\delta \Sigma}{\delta G} 
\EC \label{eq:DiffEq}
\end{align}%
\label{eq:MBEqs}%
\end{subequations}%
where $\cdot$ represents a matrix product and $\circ$ a
suitable contraction of indices
\footnote{The precise formulation can be found in
Ref.\ \onlinecite{Kugler2017a}; the indices in the functional
derivative are
$\delta \Sigma_{x\p,x} / \delta G_{y,y\p} =
-I_{t; x\p,y\p;x,y} = I_{a; x\p,y\p;y,x}$.}.
The first equation is the well-known Dyson equation, the second one
the Schwinger-Dyson equation (SDE, or equation of motion) for the self-energy,
and the last one a Bethe-Salpeter equation (BSE), where the irreducible
vertex $I_t$ is obtained by a functional derivative of $\Sigma$ w.r.t.\ $G$.
These equations together with further equations discussed below are illustrated in \FR{fig:diagram_rel}.
%

%
\begin{figure*}[t]
\includegraphics[width=1\textwidth]{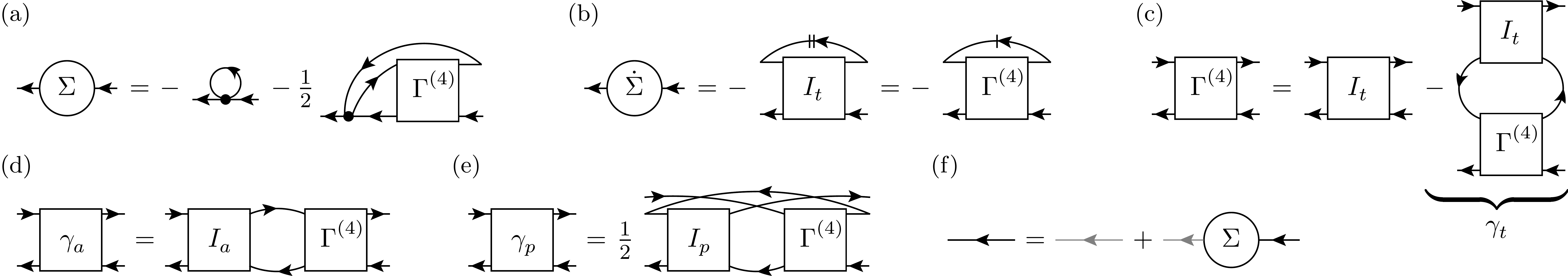}
\caption{%
Graphical representation of many-body relations, where solid lines represent dressed propagators $G$ and dots represent bare four-point vertices $\Gamma_0^{(4)}$.
(a) Schwinger-Dyson equation \ERn{eq:SchwingerDyson} for the self-energy.
(b) To perform the functional derivative $\delta \Sigma / \delta G$ in \ER{eq:DiffEq}, one sums all copies of diagrams where one $G$ line is removed. Conversely, the self-energy differentiated w.r.t.\ a scalar parameter (see main text), $\dot{\Sigma}$, is obtained by contracting [cf.\ \ER{eq:sigma_dot_I}] the vertex $I_t$ with $\dot{G}$ (line with double dash) or [cf.\ \ER{eq:sigma_dot_Gamma}] the full vertex $\Gamma^{(4)}$ with the singled-scale propagator $S$ [cf.\ \ER{eq:singlescale}, line with one dash].
(c) $\Gamma^{(4)}$ deduced from the Bethe-Salpeter equation (BSE) in the transverse channel \ERn{eq:DiffEq}. (d)--(e)
BSEs \ERn{eq:BetheSalpeter} for the reducible vertices in (d) the antiparallel channel and (e) the parallel channel.
(f) Dyson equation \ERn{eq:dyson} involving the bare propagator $G_0$ (gray line).
Note that the relations (a)--(c) suffice to generate all skeleton diagrams for the self-energy and the vertex (with all signs and prefactors written explicitly). Relations (c)--(e) together with \ER{eq:parquet} enable the parquet decomposition of the four-point vertex. Finally, the Dyson equation (f) makes the connection between bare and skeleton diagrams.%
}
\label{fig:diagram_rel}
\end{figure*}
The relation between $I_t$ and $\Sigma$
is closely related \cite{Kugler2018} to an exact
flow equation of the functional renormalization group (fRG) framework
\cite{Metzner2012,Kopietz2010}.
There, the theory evolves under the RG flow by variation
of a scale parameter $\Lambda$, introduced in the bare propagator.
Consequently, all vertex functions develop a scale dependence
(which is suppressed in the notation), and an important
role is attached to the so-called single-scale propagator
\begin{align}
S = \dot{G} - G \cdot \dot{\Sigma} \cdot G
= ( \mathbbm{1} + G \cdot \Sigma ) \cdot 
\dot{G}_0 \cdot 
( \Sigma \cdot G + \mathbbm{1} )
\EC
\label{eq:singlescale}
\end{align}
where $\dot{G}=\partial_{\Lambda} G $, etc.
If the variation of $G$ in \ER{eq:DiffEq}
is realized by varying $\Lambda$, one obtains
by inserting \ER{eq:singlescale}
\begin{subequations}
\label{eq:sigma_dot}
\begin{align}%
\dot{\Sigma} & = - I_{t} \circ \dot{G} 
= 
- I_{t} \circ ( S + G \cdot \dot{\Sigma} \cdot G )
\label{eq:sigma_dot_I} \\
& = - ( I_{t} - I_{t} \circ G \circ G \circ I_{t} + \dots) S
= - \Gamma^{(4)} \circ S
\ED
\label{eq:sigma_dot_Gamma}
\end{align}
\end{subequations}
The iterative insertion of $\dot{\Sigma}$ on the r.h.s.\
yields a ladder construction in the $t$ channel
that produces the full vertex $\Gamma^{(4)}$ from $I_t$
[cf.\ \ER{eq:DiffEq}]
and results in the well-known flow equation of the self-energy
\cite{Metzner2012, Kopietz2010}.
Finally, the relation between the full 
and the two-particle-irreducible vertices
is made precise by the parquet equation \cite{Bickers2004,Roulet1969}
\begin{equation}
\textstyle
\Gamma^{(4)} = 
R + \sum_r \gamma_r
\EC \quad
I_r = R + \sum_{r' \neq r} \gamma_{r'}
\ED
\label{eq:parquet}
\end{equation}
Here, $R$ is the totally irreducible vertex, 
whereas the vertices $\gamma_r$ with $r \in \{ a, p, t \}$
are reducible by cutting two \textit{antiparallel} lines,
two \textit{parallel} lines, or
two \textit{transverse} (antiparallel) lines, respectively
\footnote{
The term transverse refers to a horizontal space-time axis;
in using the terms antiparallel and parallel,
we adopt the nomenclature used by Roulet et al.\ \cite{Roulet1969}.
Equivalently, a common notation \cite{Rohringer2012,Rohringer2017, Wentzell2016}
for the channels $a, p, t$
is $ph, pp, \overline{ph}$, referring to
the (longitudinal) particle-hole, the particle-particle,
and the transverse (or vertical) particle-hole channel, respectively.
One also finds the labels $x, p, d$ in the literature \cite{Jakobs2010}, 
referring to the
so-called exchange, pairing, and direct channel, respectively.}.
\nocite{Rohringer2012,Rohringer2017,Wentzell2016,Jakobs2010} 
They are obtained from the irreducible ones via the BSEs [cf.\ \ER{eq:DiffEq} and \FsR{fig:diagram_rel}(c)--\FRn{fig:diagram_rel}(e)]
\begin{align}
\gamma_r & = \sigma_r \, I_r \circ G \circ G \circ \Gamma^{(4)}
\EC \ \ \sigma_a = 1 = - \sigma_t \EC \ \sigma_p = \tfrac{1}{2}
\ED
\label{eq:BetheSalpeter}
\end{align}
The relative minus sign in the $a$ and $t$ channel
stems from the fact that $\gamma_a$ and $\gamma_t$ are
related by exchange of fermionic legs.
Following the conventions of Bickers \cite{Bickers2004},
the factor of $1/2$ used in the $p$ channel and in \ER{eq:SchwingerDyson}
ensures that, when summing over all internal indices,
one does not overcount the effect of the two
indistinguishable (parallel) lines connected to
the antisymmetric vertices.
\section{Counting of diagrams}
\label{sec:diagr_count}
A key aspect in the technique of many-body perturbation theory
is that all quantities have (under certain conventions) a unique representation as a
sum of diagrams,
which can be obtained by following the so-called Feynman rules.
In order to \textit{count} the number of diagrams
via many-body integral equations,
we express all quantities as sums of diagrams
(i.e., we expand in the interaction) and collect all
combinations that lead to the same order in the interaction.
These combinations of different numbers of diagrams
yield the number of diagrams for the resulting object.
In fact, 
the multiplicative structure in the interaction translates 
into discrete convolutions of the individual 
numbers of diagrams.
Since the interaction vertices start at least at first
order in the interaction, the resulting equations can be solved iteratively.
As a first example, we count the number of diagrams 
in the full propagator $G$ at order $n$ in the interaction, $\N_G(n)$,
given the number of diagrams in the self-energy, $\N_{\Sigma}(n)$.
We know that the bare propagator has only one contribution, 
$\N_{G_0}(n) = \delta_{n,0}$,
and that the self-energy starts at first order, i.e.,
$\N_{\Sigma}(0) = 0$.
From Dyson's equation \ERn{eq:dyson}, we then see
that the number of diagrams in the full propagator 
can be generated iteratively via
\begin{equation}
\N_G(n) = \delta_{n,0} + \sum_{m=1}^{n} \N_{\Sigma}(m) \N_G(n-m)
\ED
\label{eq:diagr_count_dyson}
\end{equation}
\begin{figure*}[t]
\includegraphics[width=1\textwidth]{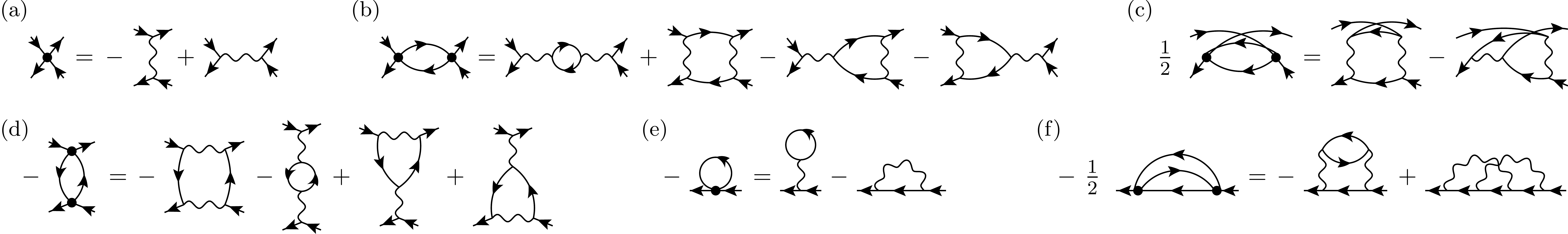}
\caption{%
Examples and translation from Hugenholtz to Feynman diagrams. 
(a) Bare (antisymmetric) four-point vertex (dot) as used for Hugenholtz diagrams expressed by direct and exchange interactions [cf.\ \ER{eq:vertex_U}, wavy lines] as used for Feynman diagrams.
(b)--(d) Diagrams for the reducible vertices $\gamma_r$ in the two-particle channels $a$, $p$, $t$, respectively. Whereas $\gamma_a$ and $\gamma_t$ have four Feynman diagrams, $\gamma_p$ has only two. In fact, inserting the direct and exchange interactions from (a) into the Hugenholtz diagram containing two equivalent propagators (parallel lines connected to antisymmetric vertices) yields only two topologically distinct diagrams, properly canceling the factor of $1/2$.
(e) First- and (f) second-order diagrams for the self-energy. The prefactor of $1/2$ is again canceled upon decomposing $\Gamma_0$.
Note that, if the electron propagators (lines) are considered as dressed ones, the above diagrams
comprise all skeleton diagrams of the four-point vertex and the self-energy up to second order.%
}
\label{fig:diagrams}
\end{figure*}
As already indicated, it is useful to define
a convolution of sequences according to
\begin{equation}
\N_1 = \N_2 \ast \N_3
\ \Leftrightarrow \
\N_1(n) = \sum_{m=0}^{n} \N_2(m) \N_3(n-m)
\ \forall n
\ED
\end{equation}
With this, we can write \ER{eq:diagr_count_dyson} in direct analogy to the 
original equation \ERn{eq:dyson} as
\begin{equation}
\N_G = \N_{G_0} + \N_{G_0} \ast \N_{\Sigma\pss} \ast \N_{G\pss}
\ED
\label{eq:diagr_count_dyson2}
\end{equation}
Similarly, we use the SDE \ERn{eq:SchwingerDyson}
and the number of diagrams in the bare 
vertex
$\N_{\Gamma^{(4)}_0}(n)=\delta_{n,1}$ to get
\begin{align}
\N_{\Sigma} = \N_{\Gamma^{(4)}_0} \ast \N_{G\pss}
+ \tfrac{1}{2}\,
\N_{\Gamma^{(4)}_0} \ast \N_{G\pss} \ast \N_{G\pss} \ast \N_{G\pss} \ast \N_{\Gamma^{(4)}}
\ED
\label{eq:diagr_count_sd}
\end{align}
We can ignore the extra minus signs when collecting topologically distinct
diagrams (for an example of 
many-body 
relations where the relative minus
signs do matter, see \AR{appendix}).
However, we have to 
keep track of prefactors 
of magnitude not equal to unity to
avoid double counting of diagrams \cite{Bickers2004}.
This is necessary as we use the antisymmetric bare four-point vertex
as building block for diagrams.
If one counts direct and exchange interactions separately, 
corresponding to an expansion in terms of the amplitude $U$ instead of the antisymmetric matrix $\Gamma_0$ in \ER{eq:vertex_U},
one attributes
two diagrams to the bare vertex 
[$\N_{\Gamma^{(4)}_0}(n)= 2 \delta_{n,1}$], and
the number of diagrams at each order
is magnified by $\N_X(n) \to \N_X(n) 2^n$. 
This corresponds to the translation from Hugenholtz 
to Feynman diagrams \cite{Negele2008}
and cancels the fractional prefactors (cf.\ \FR{fig:diagrams}).  
The further relations for the number of diagrams 
that follow from \ER{eq:DiffEq} close the set of equations
and will allow us to generate the exact numbers of diagrams
in all involved quantities.
The crucial point for this to work is that,
on the one hand,
as $\N_{\Gamma^{(4)}_0}(n) \propto \delta_{n,1}$,
the self-energy at order $n$ is generated
by $G$ (containing $\Sigma$) and $\Gamma^{(4)}$ up to order $n-1$
via \ER{eq:SchwingerDyson}.
On the other hand, \ER{eq:sigma_dot} [deduced from \ER{eq:DiffEq}]
relates $\dot{\Sigma}$ at order 
$n$ to $\Sigma$ at orders $1, \dots, n-1$
and $\Gamma^{(4)}$ at orders $1, \dots, n$.
Knowing $\N_{\Sigma}(n)$ from the SDE, 
we can thus infer $\N_{\Gamma^{(4)}}(n)$.
Then, the algorithm proceeds iteratively.
To use the differential equations, note that
a diagram of the propagator $G$
at order $n$ contains $2n+1$ lines,
and a diagram of an $m$-point vertex $\Gamma^{(m)}$ 
(we use $\Sigma = \Gamma^{(2)}$ as in Ref.\ \onlinecite{Kopietz2010})
has $(4n-m)/2$ lines.
According to the product rule, the number of differentiated
diagrams is thus given by
\begin{subequations}
\begin{align}
\N_{\dot{G}}(n) & = \N_G(n) (2n+1)
\EC \\
\N_{\dot{\Gamma}^{(m)}}(n) & = \N_{\Gamma^{(m)}}(n) (2n-\tfrac{m}{2})
\label{eq:diagr_count_gamma_dot}
\ED
\end{align}
\end{subequations}
Further, \ER{eq:sigma_dot} is easily translated into
\begin{subequations}
\label{eq:diagr_count_sigma_diff}
\begin{align}
\N_{\dot{\Sigma}} & = \N_{\Gamma^{(4)}} \ast \N_{S}
\label{eq:diagr_count_sigma_diff_a}
\\ & = \N_{I_{t}} \ast \N_{\dot{G}}
\label{eq:diagr_count_sigma_diff_b}
\end{align}
\end{subequations}
and can be transformed to give an
equation for the number of diagrams in 
the vertices $\Gamma^{(4)}$ and $I_t$. From
\ER{eq:diagr_count_sigma_diff_a}, we get
\begin{align}
\N_{\Gamma^{(4)}}(n) & = 
\Big[ \N_{\dot{\Sigma}}(n) 
- \sum_{m=1}^{n-1} \N_{\Gamma^{(4)}}(m) \N_S(n-m) 
\Big] / \N_S(0)
\EC
\label{eq:diagr_count_Gamma_from_Sigma}
\end{align}
where the number of diagrams in the single-scale propagator $S$
can be obtained from the equivalent relations%
\begin{subequations}
\begin{align}
\N_{S\pd} 
& = 
\N_{\dot{G}} - \N_{G\pd} \ast N_{\dot{\Sigma}} \ast \N_{G\pd}
\\
& = 
( \N_{\mathbbm{1}\pd} + \N_{G\pd} \ast \N_{\Sigma\pd} ) \ast \N_{\dot{G}_0} \ast
( \N_{\mathbbm{1}\pd} + \N_{\Sigma\pd} \ast \N_{G\pd} )
\EC
\end{align}
\end{subequations}
with
$\N_{\dot{G}_0}(n) = \delta_{n,0} = \N_{\mathbbm{1}}(n)$.
If we alternatively use \ER{eq:diagr_count_sigma_diff_b}
[combined with \ER{eq:DiffEq}],
we have
\begin{subequations}
\label{eq:diagr_count_It_Gamma}
\begin{align}
\N_{I_t}(n) & = 
\Big[ \N_{\dot{\Sigma}}(n) 
- \sum_{m=1}^{n-1} \N_{I_t}(m) \N_{\dot{G}}(n-m) 
\Big] / \N_{\dot{G}}(0)
\EC 
\label{eq:diagr_count_It_from_Sigma}
\\
\N_{\Gamma^{(4)}}(n) & = \N_{I_t}(n) + 
\sum_{m=1}^{n-1} \N_{\Gamma^{(4)}}(m) 
\nonumber \\
& 
\qquad \qquad \quad \times
\big( \N_G \ast \N_G \ast \N_{I_t} \big) (n-m)
\ED
\end{align}
\end{subequations}
In an analogous fashion, one can also derive the number of diagrams
in the 1PI six-point vertex $\Gamma^{(6)}$ from the exact fRG flow equation
\cite{Metzner2012, Kopietz2010} of the four-point vertex $\Gamma^{(4)}$,
\begin{align}
\N_{\dot{\Gamma}^{(4)}} & = 5\, \N_{\Gamma^{(4)}} \ast 
\N_{G\pss} \ast \N_{S\pss} \ast 
\N_{\Gamma^{(4)}}
+ \N_{\Gamma^{(6)}} \ast \N_{S\pss}
\EC
\label{eq:diagr_count_Gamma_dot}
\end{align}
together with \ER{eq:diagr_count_gamma_dot}.
A further relation is given by the SDE for $\Gamma^{(4)}$ \cite{Veschgini2013}
($\N_{\Pi} = \N_G \ast \N_G$)
\begin{align}
\N_{\Gamma^{(4)}} & = 
\N_{\Gamma^{(4)}_{0}} 
+ \tfrac{5}{2}\, \N_{\Gamma^{(4)}_{0}} \ast \N_{\Pi\pss} \ast \N_{\Gamma^{(4)}}
\nonumber \\ & \ 
+ 4\, \N_{\Gamma^{(4)}_{0}} \ast \N_{\Pi\pss} \ast \N_{\Pi\pss} 
\ast \N_{\Gamma^{(4)}} \ast \N_{\Gamma^{(4)}}
\nonumber \\ & \ 
+ \tfrac{1}{2}\, \N_{\Gamma^{(4)}_{0}} \ast \N_{G\pss} \ast \N_{\Pi\pss} \ast \N_{\Gamma^{(6)}}
\ED
\label{eq:diagr_count_Gamma_sde}
\end{align}
Finally, the number of diagrams in 
the vertex $\Gamma^{(4)}$ can be decomposed into
two-particle channels according
to the parquet equations
\ERn{eq:parquet}, \ERn{eq:BetheSalpeter}. 
By symmetry, we have $\N_{\gamma_a} = \N_{\gamma_t}$
and obtain
\begin{subequations}
\label{eq:diagr_count_parquet}
\begin{align}
\N_{\Gamma^{(4)}} 
& = 
\N_{R} + 2\, \N_{\gamma_a} + \N_{\gamma_p}
\EC
\label{eq:diagr_count_R}
\\
\N_{\gamma_r} 
& = 
|\sigma_r| ( \N_{\Gamma^{(4)}} - \N_{\gamma_r} ) \ast \N_G \ast \N_G \ast \N_{\Gamma^{(4)}}
\ED
\label{eq:diagr_count_gamma} 
\end{align}
\end{subequations}
Given $\N_{\Gamma^{(4)}}$, one can first deduce
$\N_{\gamma_r}$ and then $\N_R$.
If, conversely, the number of diagrams in the totally irreducible vertex $R$
[with $\N_{R}(0) = 0$] is fixed, 
as is the case in parquet approximations,
one can combine these equations with
\EsR{eq:diagr_count_dyson2} and \ERn{eq:diagr_count_sd}
to generate all numbers of diagrams
without the need to use the differential equations 
\ERn{eq:diagr_count_sigma_diff}.
\begin{table}[t]
\begin{tabular*}{0.48\textwidth}{@{\extracolsep{\fill}} l c c c c c c }
\hline\hline
\rule{0pt}{3ex}
$n$ & 1 & 2 & 3 & 4 & 5 & 6
\\
\\[-3.0ex]
\hline
\rule{0pt}{3ex}%
$\N_{\Gamma^{(6)}}$ & 0 & 0 & 21 & 319$\tfrac{1}{2}$ & 4180$\tfrac{1}{2}$ & 53612$\tfrac{1}{4}$
\\ 
$\N_{\Gamma^{(4)}}$ & 1 & 2$\tfrac{1}{2}$ & 15$\tfrac{1}{4}$ & 112$\tfrac{1}{8}$ & 935$\tfrac{1}{16}$ & 8630$\tfrac{5}{32}$
\\ 
$\N_{\gamma_a}$ & 0 & 1 & 6 & 42$\tfrac{1}{4}$ & 332 & 2854$\tfrac{9}{16}$
\\ 
$\N_{\gamma_p}$ & 0 & $\tfrac{1}{2}$ & 3$\tfrac{1}{4}$ & 23$\tfrac{5}{8}$ & 188$\tfrac{1}{16}$ & 1622$\tfrac{17}{32}$
\\ 
$\N_{R}$ & 1 & 0 & 0 & 4 & 83 & 1298$\tfrac{1}{2}$
\\ 
$\N_{\Sigma}$ & 1 & 1$\tfrac{1}{2}$ & 5$\tfrac{1}{4}$ & 25$\tfrac{7}{8}$ & 158$\tfrac{1}{16}$ & 1132$\tfrac{19}{32}$
\\ 
$\N_{G}$ & 1 & 2$\tfrac{1}{2}$ & 9$\tfrac{1}{4}$ & 44$\tfrac{1}{8}$ & 255$\tfrac{1}{16}$ & 1725$\tfrac{5}{32}$
\\
\\[-3.0ex]
\hline\hline
\end{tabular*}
\caption{Exact number of Hugenholtz diagrams for various vertex functions and the propagator up to interaction order 6.
The number of Feynman diagrams is obtained by $\N_X(n) \to \N_X(n) 2^n$, which cancels all fractional parts (cf.\ \FR{fig:diagrams}).%
}
\label{tab:num_diagr}
\end{table}
\section{Results}
\label{sec:results}
\subsection{Bare diagrams}
With the equations stated above, we can construct
the exact number of diagrams of the general many-body problem
for all involved quantities.
Table \ref{tab:num_diagr} shows the number of diagrams
in the different vertices, the self-energy, and the propagator up to
order 6. 
After translation from the number of Hugenholtz to Feynman diagrams
by $\N_X(n) \to \N_X(n)2^n$, $\N_G$ reproduces the numbers
already given in Ref.\ \onlinecite{Cvitanovic1978} (their Table I, first column)
and Ref.\ \onlinecite{Jishi2013} [their Eq.\ (9.10)].
\subsection{Skeleton diagrams}
For many purposes, it is convenient to work with
skeleton diagrams, i.e., diagrams in which 
all electron propagators are fully dressed ones.
Then, the bare propagator 
[with $\N_{G_0}(n) = \delta_{n,0} = \N_{\dot{G}_0}(n)$]
is replaced as building block for diagrams
by the full propagator,
for which we now use
$\N_G(n) = \delta_{n,0} = \N_{\dot{G}}(n)$.
We can directly apply the previous methods
by using those equations that are phrased with dressed propagators,
such as \EsR{eq:diagr_count_sd}, \ERn{eq:diagr_count_It_Gamma},
and \ERn{eq:diagr_count_parquet}.
Moreover, the numbers of bare and skeleton diagrams are
directly related. According to the number of lines in
an $n^{\textrm{th}}$-order diagram of an $m$-point vertex
[cf.\ \ER{eq:diagr_count_gamma_dot}], one has
\begin{align}
\N_{\Gamma^{(m)}}(n) & = 
\sum_{k=1}^{n} \N_{\Gamma^{(m)}}^{\textrm{sk}}(k) 
\big( \underbrace{\N_G \!\ast\!\cdots\!\ast\! \N_ G}_{2k-m/2} \big) (n-k)
\label{eq:diagr_count_bare_sk}
\end{align}
and can transform the number of
skeleton diagrams $\N_{\Gamma^{(m)}}^{\textrm{sk}}$ 
to bare diagrams $\N_{\Gamma^{(m)}}$.
For this, the numbers of bare diagrams in $\Sigma$ and $G$
are built up side by side, using \ER{eq:diagr_count_dyson}.
If we consider, e.g., the simplest approximation of a
finite-order \textit{skeleton} self-energy,
namely, the Hartree-Fock approximation 
with $\N_{\Sigma}^{\textrm{sk}}(n)=\delta_{n,1}$,
\ER{eq:diagr_count_bare_sk} can be used to give
$\N_{\Sigma}(n) = 0, 1, 2, 5, 14, 42, 132, \dots$
for the number of \textit{bare} self-energy diagrams.
If, conversely, the number of bare diagrams $\N_{\Gamma^{(m)}}$
is known,
we can easily construct a recursion relation
for $\N_{\Gamma^{(m)}}^{\textrm{sk}}$ by inverting \ER{eq:diagr_count_bare_sk},
\begin{align}
\N_{\Gamma^{(m)}}^{\textrm{sk}}(n) & = 
\Big[ \N_{\Gamma^{(m)}}(n) - \sum_{k=1}^{n-1} \N_{\Gamma^{(4)}}^{\textrm{sk}}(k)
\nonumber \\ & \!\!\!\!\! \times
\big( \underbrace{\N_G \!\ast\!\cdots\!\ast\! \N_ G}_{2k-m/2} \big) (n-k)
\Big] /
\big( \underbrace{\N_G \!\ast\!\cdots\!\ast\! \N_G}_{2n-m/2} \big) (0)
\ED
\raisetag{6pt}
\label{eq:diagr_count_sk_bare}
\end{align}
Table \ref{tab:num_diagr_sk} shows the
number of skeleton diagrams
in the various quantities.
The number of skeleton \textit{Feynman} diagrams
for the self-energy,
$\N^\textrm{sk}_{\Sigma}(n)2^n$,
agrees with the numbers 
given in Ref.\ \onlinecite{Molinari2006} 
[coefficients in their Eq.~(17) using $\ell=1$]
and Ref.\ \onlinecite{Zhou2008} (their Table 4.1, column 2
\footnote{Their number at order 6 should be
12018 instead of 12081.}).
\begin{table}[t]
\begin{tabular*}{0.48\textwidth}{@{\extracolsep{\fill}} l c c c c c c }
\hline\hline
\rule{0pt}{3ex}
$n$ & 1 & 2 & 3 & 4 & 5 & 6
\\
\\[-3.0ex]
\hline
\rule{0pt}{3ex}%
$\N_{\Gamma^{(6)}}^{\textrm{sk}}$ & 0 & 0 & 21 & 256$\tfrac{1}{2}$ & 2677$\tfrac{1}{2}$ & 28179$\tfrac{3}{4}$
\\ 
$\N_{\Gamma^{(4)}}^{\textrm{sk}}$ & 1 & 2$\tfrac{1}{2}$ & 10$\tfrac{1}{4}$ & 56$\tfrac{1}{8}$ & 375$\tfrac{9}{16}$ & 2931$\tfrac{21}{32}$
\\
$\N_{\gamma_a}^{\textrm{sk}}$ & 0 & 1 & 4 & 20$\tfrac{1}{4}$ & 123 & 866$\tfrac{1}{16}$
\\ 
$\N_{\gamma_p}^{\textrm{sk}}$ & 0 & $\tfrac{1}{2}$ & 2$\tfrac{1}{4}$ & 11$\tfrac{5}{8}$ & 70$\tfrac{9}{16}$ & 493$\tfrac{1}{32}$
\\ 
$\N_{R}^{\textrm{sk}}$ & 1 & 0 & 0 & 4 & 59 & 706$\tfrac{1}{2}$
\\ 
$\N_{\Sigma}^{\textrm{sk}}$ & 1 & $\tfrac{1}{2}$ & 1$\tfrac{1}{4}$ & 5$\tfrac{1}{8}$ & 28$\tfrac{1}{16}$ & 187$\tfrac{25}{32}$
\\
\\[-3.0ex]
\hline\hline
\end{tabular*}
\caption{Exact number of skeleton Hugenholtz diagrams for various vertex functions up to interaction order 6.
The number of Feynman diagrams is again obtained by $\N_X(n) \to \N_X(n) 2^n$.%
}
\label{tab:num_diagr_sk}
\end{table}
\subsection{Asymptotic behavior}
From combinatorial arguments,
it is clear that the number of diagrams exhibits
a factorial growth with the interaction order $n$.
Indeed, \FR{fig:diagrcount3} (full lines) shows the number of diagrams
in different vertex functions $\N_{\Gamma^{(m)}}$
divided by their (numerically determined) asymptote
\begin{equation}
\N_{\Gamma^{(m)}} \sim
n!n^{(m-1)/2}2^{(m-2)/2}
\EC \quad
n \gg 1
\label{eq:asymp_vertex}
\end{equation}
as a function of $1/n$. 
The fact that the curves linearly approach a finite
value demonstrates that, indeed, the correct asymptotic behavior has been identified.
We find the same proportionality factor for all vertex functions.
\begin{figure}[t]
\includegraphics[width=.48\textwidth]{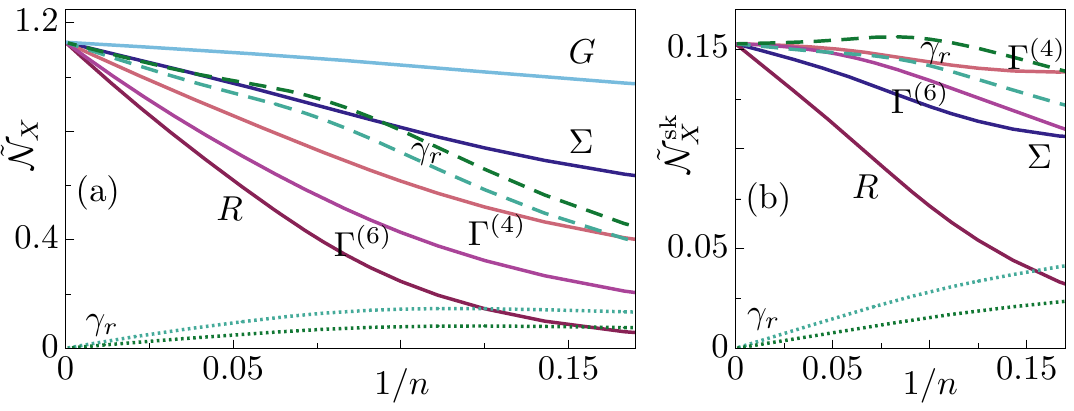}
\caption{%
Plots for the rescaled number of 
(a) bare
and (b) skeleton
diagrams with
$n$ ranging up to $1500$.
Numbers are rescaled as
$\tilde{\N}_{\Gamma^{(m)}}(n) = \N_{\Gamma^{(m)}}(n) / (n!n^{(m-1)/2}2^{(m-2)/2})$
[\ER{eq:asymp_vertex}];
$G$ is rescaled in the same way as $\Sigma=\Gamma^{(2)}$ [\ER{eq:asymp_G}];
$R$ and $\gamma_r$ ($r=a,p$, dotted) in the same way 
as $\Gamma^{(4)}$. 
Dashed lines for $\gamma_r$ account for the correct asymptote,
showing $\N_{\gamma_r}/ (4|\sigma_r|n!n^{1/2})$ [\ER{eq:asymp_gamma}].%
}
\label{fig:diagrcount3}
\end{figure}
The $m$ dependence in \ER{eq:asymp_vertex}
can be readily understood 
from the universal part of the 
exact fRG flow equations,
$\dot{\Gamma}^{(m)} = - \Gamma^{(m+2)} \circ S + \dots$
\cite{Metzner2012, Kopietz2010}.
Due to the factorial growth, we have
$\N_X(n) \gg \N_X(n-1)$ for $n \gg 1$,
and the leading behavior is determined by [using $\N_S(0)=1$ and \ER{eq:diagr_count_gamma_dot}]
\begin{equation}
\N_{\Gamma^{(m+2)}}(n) \N_{S\pss}(0) \sim
\N_{\dot{\Gamma}^{(m)}}(n) \sim 2n\, \N_{\Gamma^{(m)}}(n)
\EC
\
n \gg 1
\ED
\label{eq:asymp_diff_vertex}
\end{equation}
The asymptotes of $G$ and $\Sigma=\Gamma^{(2)}$ agree due to the 
simple relation deduced from \ER{eq:diagr_count_dyson2}
for $n \gg 1$,
\begin{equation}
\N_G(n) \sim \N_{G_0}(0) \N_{\Sigma}(n) \N_G(0)
\sim \N_{\Sigma}(n)
\sim n! n^{1/2}
\ED
\label{eq:asymp_G}
\end{equation}
The number of diagrams in the reducible vertices $\gamma_r$
divided by the same function as $\Gamma^{(4)}$ 
(dotted lines in \FR{fig:diagrcount3}) go to zero.
In fact, the correct asymptote of the reducible vertices
(as used for the dashed lines in \FR{fig:diagrcount3}) 
is found from the BSEs \ERn{eq:diagr_count_gamma}
\begin{align}
\N_{\gamma_r}(n) & \sim 2 |\sigma_r| \N_{\Gamma^{(4)}}(1)
\N_G(0) \N_G(0) \N_{\Gamma^{(4)}}(n-1) 
\nonumber \\ & \sim
4 |\sigma_r| (n-1)! n^{3/2} = 4 |\sigma_r| n! n^{1/2}
\EC \ n \gg 1
\ED
\label{eq:asymp_gamma}
\end{align}
According to \ER{eq:diagr_count_R}, 
the number of diagrams in the
totally irreducible vertex $R$
must then grow as fast as $\N_{\Gamma^{(4)}}$,
\begin{subequations}
\label{eq:asymp_R}
\begin{align}
\N_R(n) & \sim \N_{\Gamma^{(4)}}(n) \sim 2 n! n^{3/2}
\EC \\
\frac{\N_{\gamma_r}(n)}{\N_R(n)} & \sim \frac{2|\sigma_r|}{n}
\EC \quad n \gg 1
\ED
\end{align}
\end{subequations}
From \FR{fig:diagrcount3}, we indeed see that $\N_R > \N_{\gamma_a}, \N_{\gamma_p}$
for $n > 8$.
The proportionality factor of roughly $1.128$ in the asymptotics
of the bare number of diagrams
can be derived from a combinatorial approach to count
diagrams in $m$-point connected Green's function
$G^{(m)}$ (with $G=G^{(2)}$).
If the recursion relation for $G$ given in Ref.\ \onlinecite{Jishi2013} 
[their Eq.\ (9.10)]
is translated to Hugenholtz diagrams and generalized
to $m$-point functions, it reads
\begin{align}
\N_{G^{(m)}}(n) & = \frac{(2n+m/2)!}{n!4^n} - 
\sum_{k=1}^n \frac{(2k)!}{k!4^k} \N_{G^{(m)}}(n-k)
\EC
\end{align}
where the first summand accounts for all topologically distinct contractions
and the second summand removes disconnected ones.
For the asymptotic behavior, it suffices to subtract
the \textit{fully} disconnected part
[the $k=n$ summand dominates since $\N_X(n) \gg \N_X(n-1)$], and we obtain,
using $\N_{G^{(m)}}(0) = \mathit{O}(1)$ and Stirling's formula,
\begin{align}
\N_{G^{(m)}}(n) & \sim \frac{(2n+m/2)!}{n!4^n} - \frac{(2n)!}{n!4^n} 
\sim \frac{(2n)^{m/2}(2n)!}{n!4^n} 
\nonumber \\ & 
\sim \frac{2}{\sqrt{\pi}} n! n^{(m-1)/2} 2^{(m-2)/2}
\EC \ n \gg 1
\ED
\end{align}
Comparing this to \ER{eq:asymp_vertex}, we indeed find
a proportionality factor of
$2/\sqrt{\pi} \approx 1.128$
\footnote{We conclude that in Ref.\ \onlinecite{Cvitanovic1978}, Table I, first column,
the factor $C$ should read $\sqrt{2/\pi}$ instead of $\sqrt{2}/\pi$. We can numerically confirm
their coefficients $d_1$ and $d_2$ for the subleading contributions.}.
\begin{figure}[t]
\includegraphics[width=.48\textwidth]{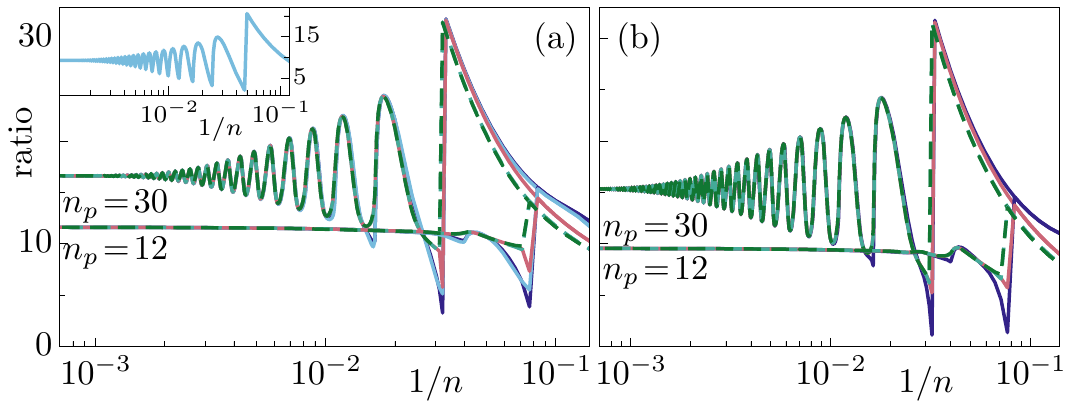}
\caption{%
Ratio of subsequent elements of (a) $\N_X$
and (b) $\N_X^{\textrm{sk}}$ in
the parquet-type approximations with $n_p = 30$ and $n_p = 12$ (see main text).
We use the same color coding as in \FR{fig:diagrcount3};
dashed lines represent $\gamma_r$.
The inset shows an analogous plot for $\N_G$, obtained
from a finite-order self-energy ($n_s=20$)
[cf.\ \ER{eq:G_Sigma_finite_order}].
The cusp for $\Gamma^{(4)}$, $\Sigma$, $G$ occurs at $1/n_p$ (inset: $1/n_s$),
and for $\gamma_r$ at $1/n_p+1$, due to the structure of the BSEs 
[cf.\ \ERn{eq:diagr_count_gamma}].%
}
\label{fig:diagrcount4}
\end{figure}
\subsection{Asymptotics of parquet approximations}
In any type of parquet approximation,
one has $\N_R(n)=0$ for $n > n_p$
(i.e., $n_p$ denotes the highest-order contribution
retained for $R$),
whereas the reducible vertices and the self-energy
still extend to arbitrarily high orders,
as determined by the self-consistent BSEs \ERn{eq:BetheSalpeter} and SDE
\ERn{eq:SchwingerDyson}.
However, in this case,
a factorial growth in the number of diagrams
[$\N_X(n) \gg \N_X(n-1)$] leading to \ER{eq:asymp_R}
would contradict a vertex $R$ of finite order.
Hence, the number of diagrams in any approximation of the parquet type
can at most grow exponentially
[$\N_X(n) / \N_X(n-1) \sim \mathit{O}(1)$].
Figure~\FRn{fig:diagrcount4} shows how the quotient of two subsequent
elements in $\N_X$ subject to (two different) parquet-type approximations
approaches a constant; it confirms
the exponential growth and reveals 
that the exponential rate 
only depends on $n_p$ for \textit{all} vertex functions.
Curiously, one finds dampened oscillations
modulating the growth
in the number of diagrams
for $n > n_p \gtrsim 10$.
An analogous phenomenon already occurs by
using the Dyson equation with a self-energy 
of finite order (cf.\ \FR{fig:diagrcount4}, inset).
Again, a factorial growth in the number of diagrams
[$\N_X(n) \gg \N_X(n-1)$]
leading to \ER{eq:asymp_G}
would contradict such an $\N_{\Sigma}$, and $\N_G$ can at most grow exponentially.
If $\N_{\Sigma}(n) = 0$ for $n > n_s$,
\ER{eq:diagr_count_dyson} is simplified to
\begin{equation}
\N_G(n) = \delta_{n,0} + \sum_{m=1}^{\textrm{min}\{n,n_s\}} \N_{\Sigma}(m) \N_G(n-m)
\ED
\label{eq:G_Sigma_finite_order}
\end{equation}
For large $n$, the factor $\N_G(n-m)$
spans over the orders $n-n_s, \dots, n$
and produces ``fading echoes'' of the abrupt fall
in the quotient which  stems from the first occurrence
of $\N_{\Sigma}(n)=0$ at $n=n_s+1$.
Even if only the skeleton diagrams
of, e.g., $\Sigma$ or $R$ are of finite order,
the resulting numbers of bare diagrams can grow at most exponentially.
The reasoning is similar:
A factorial growth in the number of diagrams [$\N_X(n) \gg \N_X(n-1)$]
would imply
$\N_{\Gamma^{(m)}}(n) \sim \N_{\Gamma^{(m)}}^{\textrm{sk}}(n_{\textrm{min}}) 
\N_G(n-n_{\textrm{min}})$, using \ER{eq:diagr_count_bare_sk} and $\N_G(0)=1$.
For $\Sigma$, one has $n_{\textrm{min}}=1$, and the result would directly
contradict \ER{eq:asymp_G}. For $R$, one has $n_{\textrm{min}}=4$
and would find a contradiction using \EsR{eq:asymp_diff_vertex}, \ERn{eq:asymp_G}, and \ERn{eq:asymp_R}.
We conclude that for any of the typical 
diagrammatic resummation approaches, one generates 
numbers of (bare) diagrams that grow at most exponentially with
interaction order $n$.
%


%
\subsection{Hubbard model}
\label{sec:Hubbard}
\begin{table}[t]
\begin{tabular*}{0.48\textwidth}{@{\extracolsep{\fill}} l r r r r r r r }
\hline\hline
\rule{0pt}{3ex}
$n$ & 1 & 2 & 3 & 4 & 5 & 6 & 7
\\
\\[-3.0ex]
\hline
\rule{0pt}{3ex}%
$\N_{\Sigma}$ & 1 & 2 & 8 & 44 & 296 & 2312 & 20384
\\
\rule{0pt}{3ex}%
$\N_{\Gamma^{(4)}}^{\uparrow\downarrow}$ & 1 & 2 & 13 & 104 & 940 & 9352 & 101080 
\\
$\N_{\gamma_a}^{\uparrow\downarrow}$ & 0 & 1 & 5 & 36 & 300 & 2760 & 27544 
\\
$\N_{\gamma_t}^{\uparrow\downarrow}$ & 0 & 0 & 3 & 30 & 282 & 2758 & 28526 
\\
$\N_{R}^{\uparrow\downarrow}$ & 1 & 0 & 0 & 2 & 58 & 1074 & 17466 
\\
\rule{0pt}{3ex}%
$\N_{\Gamma^{(4)}}^{\uparrow\uparrow}$ & 0 & 2 & 12 & 94 & 848 & 8468 & 92016
\\
$\N_{\gamma_a}^{\uparrow\uparrow}$ & 0 & 1 & 6 & 44 & 366 & 3354 & 33334 
\\
$\N_{\gamma_p}^{\uparrow\uparrow}$ & 0 & 0 & 0 & 2 & 28 & 320 & 3532 
\\
$\N_{R}^{\uparrow\uparrow}$ & 0 & 0 & 0 & 4 & 88 & 1440 & 21816 
\\
\rule{0pt}{3ex}%
$\N_{\Gamma^{(6)}}^{\uparrow\downarrow\uparrow}$ & 0 & 0 & 8 & 144 & 2072 & 28744 & 402736 
\\
$\N_{\Gamma^{(6)}}^{\uparrow\uparrow\uparrow}$ & 0 & 0 & 12 & 144 & 1872 & 25176 & 349812
\\
\\[-3.0ex]
\hline\hline
\end{tabular*}
\caption{%
Exact number of spin-resolved bare diagrams in the Hubbard model.
By symmetry, we have 
$\N_{\gamma_a}^{\uparrow\uparrow} = \N_{\gamma_t}^{\uparrow\uparrow}$,
and one further finds
$\N_{\gamma_a}^{\uparrow\downarrow} = \N_{\gamma_p}^{\uparrow\downarrow}$
[cf.\ \FR{fig:sp_diagrams} and \ER{eq:parquet_sp}].%
}
\label{tab:num_diagr_H}
\end{table}
The Hubbard model \cite{Montorsi1992} is of special interest in condensed matter physics.
In terms of diagrams, a simplification arises due to the
SU(2) spin symmetry of the model with the restrictive bare vertex ($\sigma \in \{\uparrow, \downarrow\}$)
\begin{equation}
\Gamma^{(4)}_{0;x_1\p,x_2\p;x_1,x_2} \propto
(\delta_{\sigma_1\p,\sigma_1\pp} \delta_{\sigma_2\p,\sigma_2\pp} 
-\delta_{\sigma_1\p,\sigma_2\pp} \delta_{\sigma_2\p,\sigma_1\pp} )
\, \delta_{\sigma_1\pp,\bar{\sigma}_2\pp}
\EC
\end{equation}
where $\bar{\uparrow}=\downarrow$, $\bar{\downarrow}=\uparrow$.
In this case, one can individually
count diagrams with specific spin configuration.
In other words, one can explicitly perform the spin sums
in all diagrams and actually count
only those diagrams that do not vanish
under the spin restriction.
\begin{table}[t]
\begin{tabular*}{0.48\textwidth}{@{\extracolsep{\fill}} l r r r r r r r }
\hline\hline
\rule{0pt}{3ex}
$n$ & 1 & 2 & 3 & 4 & 5 & 6 & 7
\\
\\[-3.0ex]
\hline
\rule{0pt}{3ex}%
$\N_{\Sigma}^{\textrm{sk}}$ & 1 & 1 & 2 & 9 & 54 & 390 & 3268
\\
\rule{0pt}{3ex}%
$\N_{\Gamma^{(4)}}^{\textrm{sk}\uparrow\downarrow}$ & 1 & 2 & 9 & 54 & 390 & 3268 & 30905
\\
$\N_{\gamma_a}^{\textrm{sk}\uparrow\downarrow}$ & 0 & 1 & 3 & 17 & 112 & 850 & 7289
\\
$\N_{\gamma_t}^{\textrm{sk}\uparrow\downarrow}$ & 0 & 0 & 3 & 18 & 120 & 928 & 8029
\\
$\N_{R}^{\textrm{sk}\uparrow\downarrow}$ & 1 & 0 & 0 & 2 & 46 & 640 & 8298
\\
\rule{0pt}{3ex}%
$\N_{\Gamma^{(4)}}^{\textrm{sk}\uparrow\uparrow}$ & 0 & 2 & 8 & 48 & 352 & 2978 & 28376
\\
$\N_{\gamma_a}^{\textrm{sk}\uparrow\uparrow}$ & 0 & 1 & 4 & 21 & 136 & 1028 & 8768
\\
$\N_{\gamma_p}^{\textrm{sk}\uparrow\uparrow}$ & 0 & 0 & 0 & 2 & 16 & 126 & 1064
\\
$\N_{R}^{\textrm{sk}\uparrow\uparrow}$ & 0 & 0 & 0 & 4 & 64 & 796 & 9776
\\
\rule{0pt}{3ex}%
$\N_{\Gamma^{(6)}}^{\textrm{sk}\uparrow\downarrow\uparrow}$ & 0 &0 & 8 & 120 & 1376 & 15648 & 185296
\\
$\N_{\Gamma^{(6)}}^{\textrm{sk}\uparrow\uparrow\uparrow}$ & 0 &0 & 12 & 108 & 1188 & 13464 & 160236
\\
\\[-3.0ex]
\hline\hline
\end{tabular*}
\caption{%
Exact number of spin-resolved skeleton diagrams in the Hubbard model,
where we again have
$\N_{\gamma_a}^{\textrm{sk}\uparrow\uparrow} = \N_{\gamma_t}^{\textrm{sk}\uparrow\uparrow}$
and 
$\N_{\gamma_a}^{\textrm{sk}\uparrow\downarrow} = \N_{\gamma_p}^{\textrm{sk}\uparrow\downarrow}$.%
}
\label{tab:num_diagr_Hsk}
\end{table}
So far, we have considered diagrams 
that contain summations over \textit{all}
internal degrees of freedom---including spin.
Generally, our algorithm cannot give the functional dependence
of the diagrams and, in particular, does not
give the spin dependence of the diagrams.
If one writes the relations stated above
with their explicit spin dependence (as done in \AR{appendix}),
one finds that the SDE relates the self-energy to
the vertex with different spins at the external legs.
However, the differential equations contain
a summation over all spin configurations of the vertex.
Thus, \EsR{eq:diagr_count_Gamma_from_Sigma} and
\ERn{eq:diagr_count_It_from_Sigma}
cannot be used to deduce the 
number of spin-resolved vertex diagrams.
As already mentioned, for \textit{approximate} many-body approaches that 
do allow for an iterative construction,
such as parquet-type approximations,
we need not make use of the differential equations.
We could therefore easily construct the corresponding
numbers of spin-resolved diagrams.
However, here, we prefer to 
give low-order results for
the \textit{exact} numbers of diagrams
for all the different vertex functions
by resorting to known results:
We use exact numbers of diagrams
for a specific quantity not considered in this work, which are
obtained by Monte Carlo sampling up to order 7 
in Ref.\ \onlinecite{VanHoucke2016} (their Table I).
From this, we can deduce the number of diagrams in 
the totally irreducible vertex $R$ and, then, generate the numbers for 
all further vertex functions studied here.
Using spin symmetry, only a few spin configurations of the vertices
are actually relevant:
One-particle properties must be independent of spin;
for two- and three-particle vertices, it suffices to consider those with
identical spins and those with two different pairs of spins.
In \AR{appendix}, we explain the labeling and give further relations
that follow from the SU(2) spin symmetry and rely on cancelations
of diagrams. 
Table \ref{tab:num_diagr_H} gives the exact number of bare
diagrams for the Hubbard model up to order 7;
Table \ref{tab:num_diagr_Hsk} gives the corresponding numbers of
skeleton diagrams. 
The numbers for $\N_{\Sigma}^{\textrm{sk}}$ up to order 6
agree with those of Ref.\ \onlinecite{Zhou2008} (their Table 4.1, column 3).
Note that, for spin-resolved diagrams of the Hubbard model, we can use the internal spin summations to express all Hugenholtz diagrams in terms of the bare vertex $\Gamma_0^{\uparrow\downarrow}$ with fixed spins, containing only one diagram. Hence, the number of spin-resolved Hugenholtz and Feynman diagrams for this model are equal (cf.\ \FR{fig:sp_diagrams}).
\begin{figure}[t]
\includegraphics[width=0.48\textwidth]{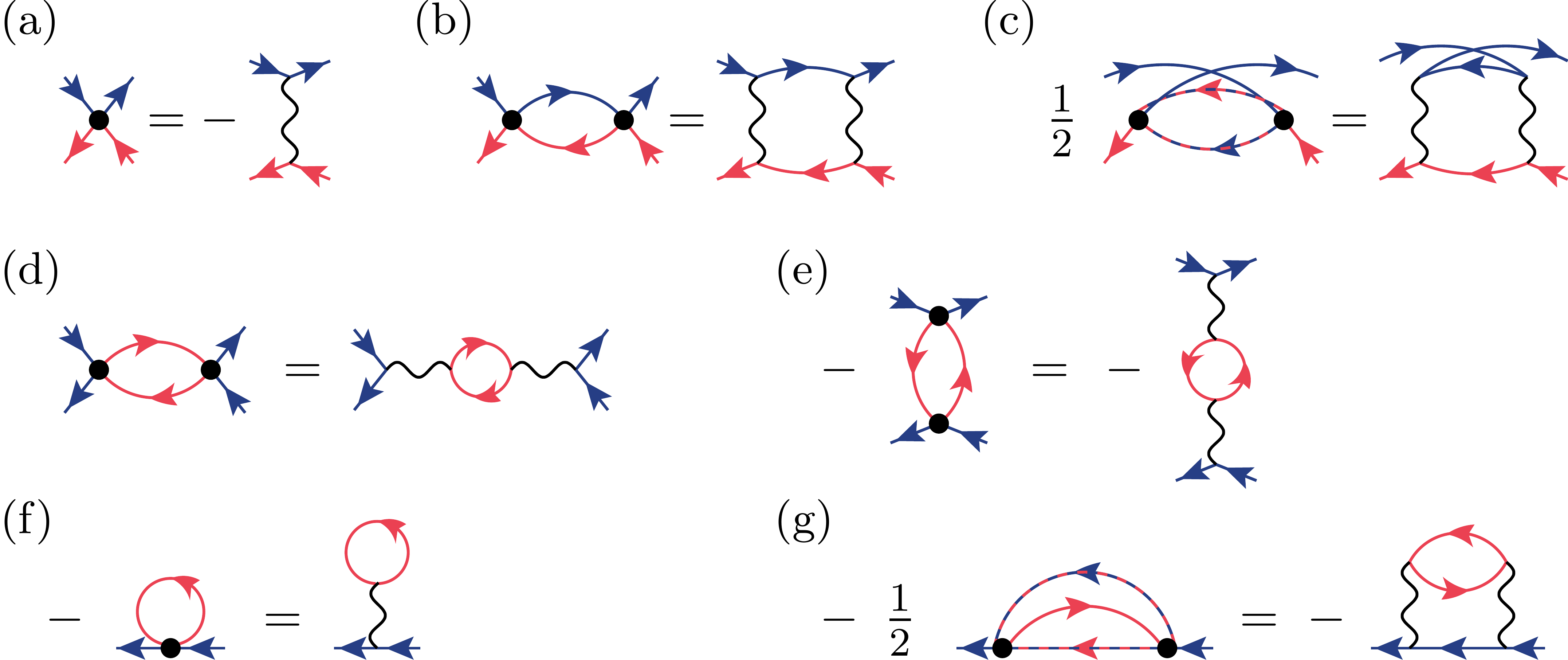}
\caption{%
Spin-resolved diagrams of the Hubbard model in the Hugenholtz and Feynman representation up to second order. Blue (dark) lines denote spin-up and red (light) lines spin-down propagators; dashed lines symbolize a sum over spin.
Panels (a)--(c) give diagrams for $\Gamma_0^{\uparrow\downarrow}$, $\gamma_a^{\uparrow\downarrow}$, and $\gamma_p^{\uparrow\downarrow}$; (d)--(e) for $\gamma_a^{\uparrow\uparrow}$, and $\gamma_t^{\uparrow\uparrow}$; (f)--(g) for $\Sigma$.
Viewed with full propagators, these are all skeleton diagrams entering $\Gamma^{(4)}$ and $\Sigma$ up to second order. We explicitly see that the numbers of Hugenholtz and Feynman diagrams are equal.%
}
\label{fig:sp_diagrams}
\end{figure}
It is interesting to compare the
number of diagrams in the four-point vertex
with identical and different spins.
On top of the numbers given in Tables
\ref{tab:num_diagr_H} and \ref{tab:num_diagr_Hsk},
our algorithm can also determine the
asymptotic behavior of, e.g., the relation between
$\N_{\Gamma^{(4)}}^{\uparrow\uparrow}$ and
$\N_{\Gamma^{(4)}}^{\uparrow\downarrow}$.
If we consider skeleton diagrams, the
SDE \ERn{eq:H_SDE_Sigma} with $\N_{\Gamma^{(4)}_0}^{\textrm{sk}\uparrow\downarrow}(n)=\delta_{n,1}$ yields
$\N_{\Sigma}^\textrm{sk}(n+1) = \N_{\Gamma^{(4)}}^{\textrm{sk}\uparrow\downarrow}(n)$.
Combined with the (super) factorial growth of $\N_{\Sigma}^\textrm{sk}$, this gives
\begin{equation}
n \N_{\Sigma}^\textrm{sk}(n) \gtrsim \N_{\Sigma}^\textrm{sk}(n+1) = \N_{\Gamma^{(4)}}^{\textrm{sk}\uparrow\downarrow}(n)
\EC \quad n \gg 1
\ED
\end{equation}
On the other hand, \ER{eq:diagr_count_gamma_dot} and \ER{eq:H_diff_Sigma_G}
together with the knowledge that $\N_R$ asymptotically dominates $\N_{\Gamma^{(4)}}$
can be used to obtain
\begin{align}
2n \N_{\Sigma}^\textrm{sk}(n) 
& \lesssim 
\N_{I_t}^{\textrm{sk}\uparrow\uparrow}(n) + \N_{I_t}^{\textrm{sk}\uparrow\downarrow}(n)
\nonumber \\
& \lesssim 
\N_{\Gamma^{(4)}}^{\textrm{sk}\uparrow\uparrow}(n) + \N_{\Gamma^{(4)}}^{\textrm{sk}\uparrow\downarrow}(n)
\EC \quad n \gg 1
\ED
\end{align}
Dividing both equations, we find that, according to
\begin{equation}
\N_{\Gamma^{(4)}}^{\uparrow\uparrow}(n)/\N_{\Gamma^{(4)}}^{\uparrow\downarrow}(n)
\sim
\N_{\Gamma^{(4)}}^{\textrm{sk}\uparrow\uparrow}(n)/\N_{\Gamma^{(4)}}^{\textrm{sk}\uparrow\downarrow}(n) \gtrsim 1
\EC \quad n \gg 1
\EC
\end{equation}
the number of diagrams for the effective interaction between same
spins asymptotically approaches the one between different spins from above for
large interaction orders.
\section{Conclusion}
\label{sec:conclusions}
We have presented an iterative
algorithm to count the number of Feynman diagrams 
inherent in many-body integral equations.
We have used it to count the
exact number of bare and skeleton diagrams
in various vertex function and different two-particle
channels. 
Our algorithm can easily be applied to many-body relations of different
forms and levels of approximation,
such as the parquet formalism
\cite{Bickers2004, Roulet1969} and
its simplified variant FLEX \cite{Bickers2004},
other approaches based on
Hedin's equations \cite{Hedin1965, Molinari2006} including the famous GW approximation 
\cite{Aryasetiawan1998, Molinari2005},
$\Phi$-derivable results deduced from a specific
approximation of the Luttinger-Ward functional 
\cite{Bickers2004, Luttinger1960, Baym1962},
and truncated flows of the
functional renormalization group \cite{Metzner2012, Kopietz2010, Kugler2017, Kugler2017a}.
Due to its iterative structure, the algorithm
allows us to numerically access arbitrarily large interaction orders
and gain analytical insight
into the asymptotic behavior.
First, we have extracted a leading dependence
of $n!n^{(m-1)/2}2^{(m-2)/2}$ in the number of diagrams
of an $m$-point 1PI vertex.
Second, we have shown that the number of diagrams in the totally irreducible 
four-point vertex exceeds those of the reducible ones for 
interaction orders $n > 8$ and asymptotically
contains \textit{all} diagrams of the four-point vertex
[i.e., $\N_{\gamma_r}(n)/\N_R(n) \to 0$ as $n \to \infty$].
Third, we have argued that any of the typical diagrammatic
resummation procedures, including any
type of parquet approximation,
can support an exponential growth only in the number of diagrams.
This is in contrast to the factorial growth in the exact number of diagrams.
It is therefore likely that the corresponding 
approximate series expansions do have a finite radius of convergence.
We believe that the techniques and results presented in this paper
will be useful for various applications of Green's-functions
methods as well as approaches that directly sum diagrams, such as
finite-order approximations or diagrammatic
Monte Carlo \cite{VanHoucke2010}.

\begin{acknowledgments}
The author wishes to thank E.\ Kozik, D.\ Schimmel, J.\ von Delft, and F.\ Werner for useful discussions.
Support by the Cluster of Excellence
Nanosystems Initiative Munich and funding from
the research school IMPRS-QST is acknowledged.
\end{acknowledgments}
\appendix

\section{Relations for the Hubbard model}
\label{appendix}
The spin symmetry in the Hubbard model
allows us to focus on a small set of vertex functions
when counting diagrams.
By spin conservation, an $n$-particle vertex depends
on only $n$ spins.
Using the $\mathbb{Z}_2$ symmetry, it is clear
that self-energy diagrams do not depend on spin,
while, for the four-point vertex, it suffices
to consider
\begin{align}
\N_{\Gamma^{(4)}}^{\uparrow\uparrow}
:=
\N_{\Gamma^{(4)}}^{\uparrow\uparrow;\uparrow\uparrow}
\EC \quad
\N_{\Gamma^{(4)}}^{\uparrow\downarrow}
:=
\N_{\Gamma^{(4)}}^{\uparrow\downarrow;\uparrow\downarrow}
\ED
\end{align}
Here, we write the spin indices  of the vertex
in the order of \ER{eq:general_theory}
as superscripts of $\N$.
The classification of four-point diagrams into two-particle channels
depends on the labels of the external legs. 
By crossing symmetry, we have 
$\N_{\gamma_a}^{\uparrow\uparrow}=\N_{\gamma_t}^{\uparrow\uparrow}$
and find for different spins
\begin{subequations}
\begin{align}
\N_{\gamma_p}^{\uparrow\downarrow}
& :=
\N_{\gamma_p}^{\uparrow\downarrow;\uparrow\downarrow}
= 
\N_{\gamma_p}^{\uparrow\downarrow;\downarrow\uparrow}
\EC
\\
\N_{\gamma_a}^{\uparrow\downarrow}
& :=
\N_{\gamma_a}^{\uparrow\downarrow;\uparrow\downarrow}
= 
\N_{\gamma_t}^{\uparrow\downarrow;\downarrow\uparrow}
\EC
\\
\N_{\gamma_t}^{\uparrow\downarrow}
& :=
\N_{\gamma_t}^{\uparrow\downarrow;\uparrow\downarrow}
= 
\N_{\gamma_a}^{\uparrow\downarrow;\downarrow\uparrow}
\ED
\end{align}
\end{subequations}
For the six-point vertex, we need to consider only
(the semicolon again separates incoming and outgoing lines)
\begin{align}
\N_{\Gamma^{(6)}}^{\uparrow\uparrow\uparrow}
:=
\N_{\Gamma^{(6)}}^{\uparrow\uparrow\uparrow;\uparrow\uparrow\uparrow}
\EC \quad
\N_{\Gamma^{(6)}}^{\uparrow\downarrow\uparrow}
:=
\N_{\Gamma^{(6)}}^{\uparrow\downarrow\uparrow;\uparrow\downarrow\uparrow}
\ED
\end{align}
The SU(2) spin symmetry further relates the remaining components of the four-point vertex
by \cite{Rohringer2012}
\begin{equation}
\Gamma^{(4)}_{p\p\uparrow,q\p\uparrow;p\vpp\uparrow,q\vpp\uparrow}
=
\Gamma^{(4)}_{p\p\uparrow,q\p\downarrow;p\vpp\downarrow,q\vpp\uparrow}
-
\Gamma^{(4)}_{p\p\uparrow,q\p\downarrow;q\vpp\downarrow,p\vpp\uparrow}
\EC
\end{equation}
where we have decomposed the quantum number $x$ into
$p$ and $\sigma$.
However, this subtraction involves
cancelations of diagrams
as opposed to the summation 
of topologically distinct, independent diagrams
we have encountered so far.
This can already be seen at first order where $\N_{\Gamma^{(4)}_0}^{\uparrow\uparrow}=0$.
Such cancelations of diagrams can only
change the number of diagrams by a multiple of 2.
Consequently, we infer that
\begin{equation}
2\N_{\Gamma^{(4)}}^{\uparrow\downarrow}-\N_{\Gamma^{(4)}}^{\uparrow\uparrow}
\in 2 \mathbb{N}_0
\ED
\end{equation}
If we further invoke the channel decomposition with crossing symmetries, 
we find that  all of
\begin{equation}
2\N_R^{\uparrow\downarrow}-\N_R^{\uparrow\uparrow} 
\EC \quad
2\N_{\gamma_p}^{\uparrow\downarrow}-\N_{\gamma_p}^{\uparrow\uparrow} 
\EC \quad
\N_{\gamma_a}^{\uparrow\downarrow}+\N_{\gamma_t}^{\uparrow\downarrow}-\N_{\gamma_a}^{\uparrow\uparrow}
\end{equation}
are nonnegative, even numbers (as can explicitly be checked in Tables \ref{tab:num_diagr_H} and \ref{tab:num_diagr_Hsk}).
Next, we perform the spin summation
in the different many-body relations
stated in \SR{sec:diagr_count}.
Starting with 
\EsR{eq:diagr_count_sd} and \ERn{eq:diagr_count_sigma_diff} for the self-energy,
we get
\begin{subequations}
\begin{align}
\N_{\Sigma} & = \N_{\Gamma^{(4)}_0}^{\uparrow\downarrow} \ast \N_G + \N_{\Gamma^{(4)}_0}^{\uparrow\downarrow} \ast \N_\Pi \ast \N_G \ast \N_{\Gamma^{(4)}}^{\uparrow\downarrow}
\EC
\label{eq:H_SDE_Sigma}
\\
\N_{\dot{\Sigma}} & = (\N_{\Gamma^{(4)}}^{\uparrow\downarrow} + \N_{\Gamma^{(4)}}^{\uparrow\uparrow}) \ast \N_S 
\label{eq:H_diff_Sigma}
\\
& = (\N_{I_t}^{\uparrow\downarrow} + \N_{I_t}^{\uparrow\uparrow}) \ast \N_{\dot{G}} 
\ED
\label{eq:H_diff_Sigma_G}
\end{align}
\end{subequations}
From \EsR{eq:diagr_count_Gamma_dot} and \ERn{eq:diagr_count_Gamma_sde},
we similarly get for the four-point vertex ($\N_{\dot{\Pi}_S}=\N_G \ast \N_S$)
\begin{subequations}
\begin{align}
\N_{\dot{\Gamma}^{(4)}}^{\uparrow\downarrow} & = 2\, \N_{\Gamma^{(4)}}^{\uparrow\downarrow} \ast \N_{\dot{\Pi}_S} \ast \N_{\Gamma^{(4)}}^{\uparrow\downarrow} + 2\, \N_{\Gamma^{(4)}}^{\uparrow\downarrow} \ast \N_{\dot{\Pi}_S} \ast \N_{\Gamma^{(4)}}^{\uparrow\uparrow} \nonumber 
\\
& \ + 2\, \N_{\Gamma^{(6)}}^{\uparrow\downarrow\uparrow} \ast N_S 
\EC
\label{eq:H_diff_Gamma_1}
\\
\N_{\dot{\Gamma}^{(4)}}^{\uparrow\uparrow} & = \tfrac{5}{2}\, \N_{\Gamma^{(4)}}^{\uparrow\uparrow} \ast \N_{\dot{\Pi}_S} \ast \N_{\Gamma^{(4)}}^{\uparrow\uparrow} + 2\, \N_{\Gamma^{(4)}}^{\uparrow\downarrow} \ast \N_{\dot{\Pi}_S} \ast \N_{\Gamma^{(4)}}^{\uparrow\downarrow} \nonumber 
\\
& \ + \N_{\Gamma^{(6)}}^{\uparrow\downarrow\uparrow} \ast N_S + \N_{\Gamma^{(6)}}^{\uparrow\uparrow\uparrow} \ast N_S
\EC
\label{eq:H_diff_Gamma_2}
\\
\N_{\Gamma^{(4)}}^{\uparrow\downarrow} & = \N_{\Gamma^{(4)}_0}^{\uparrow\downarrow} + 2\, \N_{\Gamma^{(4)}_0}^{\uparrow\downarrow} \ast \N_\Pi \ast \N_{\Gamma^{(4)}}^{\uparrow\downarrow} \nonumber
\\
& \ + \N_{\Gamma^{(4)}_0}^{\uparrow\downarrow} \ast \N_\Pi \ast \N_{\Gamma^{(4)}}^{\uparrow\uparrow} + \N_{\Gamma^{(4)}_0}^{\uparrow\downarrow} \ast \N_\Pi \ast \N_{\Gamma^{(6)}}^{\uparrow\downarrow\uparrow} \nonumber
\\
& \ + 3\, \N_{\Gamma^{(4)}_0}^{\uparrow\downarrow} \ast \N_\Pi \ast \N_\Pi \ast \N_{\Gamma^{(4)}}^{\uparrow\downarrow} \ast \N_{\Gamma^{(4)}}^{\uparrow\downarrow} \nonumber
\\
& \ + 4\, \N_{\Gamma^{(4)}_0}^{\uparrow\downarrow} \ast \N_\Pi \ast \N_\Pi \ast \N_{\Gamma^{(4)}}^{\uparrow\downarrow} \ast \N_{\Gamma^{(4)}}^{\uparrow\uparrow} 
\EC
\label{eq:H_SDE_Gamma_1}
\\
\N_{\Gamma^{(4)}}^{\uparrow\uparrow} & = 2\, \N_{\Gamma^{(4)}_0}^{\uparrow\downarrow} \ast \N_\Pi \ast \N_{\Gamma^{(4)}}^{\uparrow\downarrow} \nonumber
\\
& \ + \N_{\Gamma^{(4)}_0}^{\uparrow\downarrow} \ast \N_\Pi \ast \N_{\Gamma^{(4)}}^{\uparrow\uparrow} + \N_{\Gamma^{(4)}_0}^{\uparrow\downarrow} \ast \N_\Pi \ast \N_{\Gamma^{(6)}}^{\uparrow\downarrow\uparrow} \nonumber
\\
& \ + 4\, \N_{\Gamma^{(4)}_0}^{\uparrow\downarrow} \ast \N_\Pi \ast \N_\Pi \ast \N_{\Gamma^{(4)}}^{\uparrow\downarrow} \ast \N_{\Gamma^{(4)}}^{\uparrow\downarrow} \nonumber
\\
& \ + 3\, \N_{\Gamma^{(4)}_0}^{\uparrow\downarrow} \ast \N_\Pi \ast \N_\Pi \ast \N_{\Gamma^{(4)}}^{\uparrow\downarrow} \ast \N_{\Gamma^{(4)}}^{\uparrow\uparrow}
\ED
\label{eq:H_SDE_Gamma_2}
\end{align}
\end{subequations}

Finally, we resolve the parquet equations \ERn{eq:diagr_count_parquet} in their spin configurations and obtain
\begin{subequations}
\label{eq:parquet_sp}
\begin{align}
\N_{\Gamma^{(4)}}^{\sigma\sigma'} & = \N_R^{\sigma\sigma'} + \textstyle{\sum_r} \N_{\gamma_r}^{\sigma\sigma'}
\EC
\\
\N_{I_r}^{\sigma\sigma'} & = \N_{\Gamma^{(4)}}^{\sigma\sigma'} - \N_{\gamma_r}^{\sigma\sigma'}
\EC
\\
\N_{\gamma_a}^{\uparrow\downarrow} & = \N_{I_a}^{\uparrow\downarrow} \ast \N_\Pi \ast \N_{\Gamma^{(4)}}^{\uparrow\downarrow} 
\EC
\\
\N_{\gamma_p}^{\uparrow\downarrow} & = \N_{I_p}^{\uparrow\downarrow} \ast \N_\Pi \ast \N_{\Gamma^{(4)}}^{\uparrow\downarrow} 
\EC
\\
\N_{\gamma_t}^{\uparrow\downarrow} & = \N_{I_t}^{\uparrow\downarrow} \ast \N_\Pi \ast \N_{\Gamma^{(4)}}^{\uparrow\uparrow} + \N_{I_t}^{\uparrow\uparrow} \ast \N_\Pi \ast \N_{\Gamma^{(4)}}^{\uparrow\downarrow}
\EC
\\
\N_{\gamma_a}^{\uparrow\uparrow} & = \N_{I_a}^{\uparrow\uparrow} \ast \N_\Pi \ast \N_{\Gamma^{(4)}}^{\uparrow\uparrow} + \N_{I_t}^{\uparrow\downarrow} \ast \N_\Pi \ast \N_{\Gamma^{(4)}}^{\uparrow\downarrow} 
\EC
\\
\N_{\gamma_p}^{\uparrow\uparrow} & = \tfrac{1}{2} \N_{I_p}^{\uparrow\uparrow} \ast \N_\Pi \ast \N_{\Gamma^{(4)}}^{\uparrow\uparrow} 
\EC
\\
\N_{\gamma_t}^{\uparrow\uparrow} & = \N_{I_t}^{\uparrow\uparrow} \ast \N_\Pi \ast \N_{\Gamma^{(4)}}^{\uparrow\uparrow} + \N_{I_t}^{\uparrow\downarrow} \ast \N_\Pi \ast \N_{\Gamma^{(4)}}^{\uparrow\downarrow}
\ED 
\end{align}
\end{subequations}
In \SR{sec:diagr_count}, we combined the
Schwinger-Dyson with differential (or flow)
equations to iteratively construct the exact number
of diagrams.
Here, we see that the Schwinger-Dyson equations of $\Sigma$ [\ER{eq:H_SDE_Sigma}]
and $\Gamma^{(4)}$ [\EsR{eq:H_SDE_Gamma_1} and \ERn{eq:H_SDE_Gamma_2}] 
contain the corresponding higher-point
vertex $\Gamma^{(4)}$ and $\Gamma^{(6)}$, respectively,
only in the configuration with different spins.
However, the differential equations [\EsR{eq:H_diff_Sigma} and \ERn{eq:H_diff_Sigma_G} and
\EsR{eq:H_diff_Gamma_1} and \ERn{eq:H_diff_Gamma_2}]
involve the same higher-point vertex in all of its spin configurations.
It is for this reason that one cannot iteratively construct the exact number of
\textit{spin-resolved} diagrams.
However, the equations can easily be used to generate the number of
diagrams in approximations that do allow for an iterative construction, such as parquet-type
approximations or approximations that involve a finite number of
known (bare or skeleton) diagrams.

\bibliographystyle{apsrev4-1}
\bibliography{references}
\end{document}